\documentclass[pra,twocolumn,amsmath,amssymb,superscriptaddress]{revtex4}

\usepackage{graphicx}

\def\perm{\mathop{\rm Per}}

\begin{document}

\smallskip

\title{Computational complexity of exterior products and multi-particle
amplitudes of non-interacting fermions in entangled states}

\author{Dmitri~A.~Ivanov}
\affiliation{Institute for Theoretical Physics, ETH Z\"urich,
8093 Z\"urich, Switzerland}
\affiliation{Department of Physics, University of Z\"urich, 
8057 Z\"urich, Switzerland}

\begin{abstract}
Noninteracting bosons were proposed to be used for a
demonstration of quantum-computing supremacy in a boson-sampling setup.
A similar demonstration with fermions would require that the fermions
are initially prepared in an entangled state. I suggest that
pairwise entanglement of fermions would be sufficient for this purpose.
Namely, it is shown that computing multi-particle scattering amplitudes
for fermions entangled pairwise in groups of four single-particle states 
is \#P hard. In linear algebra, such amplitudes are expressed as exterior
products of two-forms of rank two. In particular, a permanent of a $N\times N$ matrix may be
expressed as an exterior product of $N^2$ two-forms of rank two in dimension
$2N^2$, which establishes the \#P-hardness of the latter.
\end{abstract}

\maketitle

\section{Introduction}
\label{sec:Introduction}

Quantum devices are believed to have potential to outperform classical
computers \cite{nielsen-2000}. One of the challenges in the field of quantum computing
is characterizing the scope of computational tasks where quantum computers would be useful
(the most famous example of such a task is Shor's factorization
algorithm \cite{shor-1997}). On the hardware side, there exist numerous proposals of
quantum-computing devices, but a suitable scalable hardware still needs
to be invented \cite{ladd-2010}.

\begin{figure}
\centerline{\includegraphics[width=.45\textwidth]{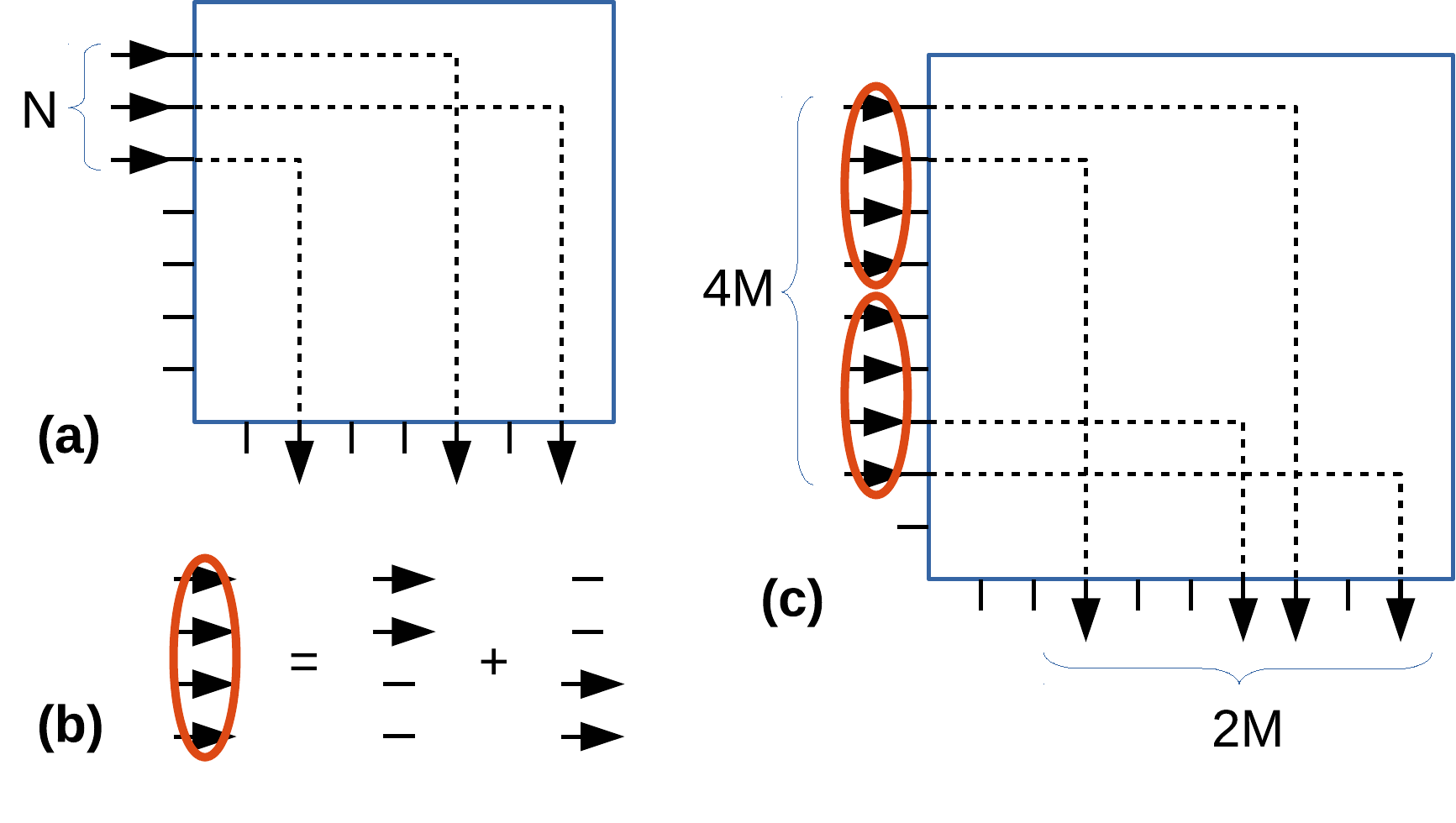}}
\caption{{\bf (a)} The Boson-Sampling setup. The square represent the scattering matrix
with rows corresponding to the input channels and columns to the output channels.
Dashed lines represent one of the multi-particle processes participating in the
interference.
{\bf (b)} The four-channel entanglement of Eq.~(\ref{four-entangled}) [the factor
$1/\sqrt{2}$ is omitted for simplicity]. {\bf (c)} The Fermion-Sampling setup with
entangled quadruplets.} 
\label{fig-scattering}
\end{figure}

One approach which targets the two above goals simultaneously is the
so-called ``quantum supremacy'' demonstration: finding a task (even
possibly useless for practical purposes) that can be
efficiently performed by a quantum device, but not by a classical
computer (see, e.g., Ref.~\onlinecite{aaronson-2016} and references
therein). An example of such a ``quantum supremacy'' task is the
Boson-Sampling proposal \cite{aaronson-2013}: noninteracting bosons are sent
to a subset of input channels of a specially designed scattering matrix
(Fig.~\ref{fig-scattering}a). After scattering, the bosons are distributed
among the output channels, with the probabilities determined by the
amplitudes of the corresponding multi-particle scattering processes.
The authors of the proposal argue that modeling this sampling
process on a classical computer would most likely require an exponentially
large computational effort (assuming the P$\ne$NP conjecture).

The key reason for the quantum supremacy of Boson-Sampling is the
computational complexity of the corresponding multi-particle amplitudes.
For noninteracting bosons, these amplitudes are given by the matrix permanent:
\begin{equation}
\perm (A) = \sum_\sigma \prod_{i=1}^N a_{i\sigma(i)}\, .
\label{perm-1}
\end{equation}
Here $A$ is a square $N\times N$ matrix with entries $a_{ij}$,
and the sum is performed over all permutations $\sigma$ of $N$ elements
(in application to the Boson-Sampling setup, $A$ is the submatrix of the
full scattering matrix spanned by the input and output channels).
The (exact) computation of a permanent is, in turn, known to be a
\#P-complete problem \cite{valiant-1979}, which is therefore believed
(assuming the P$\ne$NP conjecture) to be not solvable on classical
computers in polynomial time. Note however that, since Boson-Sampling is
not equivalent to computing a permanent, the actual argument in favor of quantum
supremacy of Boson-Sampling is more involved: see Ref.~\cite{aaronson-2013}
for details.

At the same time, the straightforward counterpart of the Boson-Sampling proposal with
fermions does not work:
the corresponding amplitudes for fermions are given by determinants, which are computable
in polynomial time \cite{papadimitriou-1994}. The resolution of this apparent
``supersymmetry breaking'' is that it is the non-Gaussian property of the initial
state that is crucial for the complexity of the quantum computation 
\cite{bravyi-2006,deMelo-2013}. For bosons, the state of one boson per channel
is non-Gaussian and therefore provides a complexity resource. For fermions, on
the other hand, the single-particle state is Gaussian, and therefore manipulations
with such states do not raise complexity beyond the single-particle level.
This difference was illustrated in Ref.~\onlinecite{shchesnovich-2015},
where it was shown that Boson-Sampling can, in fact, be simulated with fermions,
provided that the fermions are initially prepared in a specially entangled state.

The construction of Ref.~\onlinecite{shchesnovich-2015} involves fermions with a large number
of internal quantum degrees of freedom (equal to the number of particles). One can try to
optimize this construction by using simpler non-Gaussian states of fermions.
One of the simplest non-Gaussian states is the entangled state of two fermions
in four single-particle states:
\begin{equation}
\Psi_4=\frac{1}{\sqrt{2}}\left( \left| 1100 \right\rangle + \left| 0011 \right\rangle \right)\, ,
\label{four-entangled}
\end{equation}
where 1100 and 0011 refer to the fermionic occupation numbers in the four states
(Fig.~\ref{fig-scattering}b). 

The goal of the present paper is to demonstrate that, if the initial state of
fermions is given by a product of the entangled states (\ref{four-entangled}), then the
multi-particle amplitudes of a general noninteracting evolution are \#P-hard,
similarly to the Boson-Sampling proposal. Specifically, 
consider a scattering problem for $2M$ noninteracting fermions distributed over
$4M$ input channels divided into $M$ quadruplets, each of those quadruplets
being prepared in the state $\Psi_4$ (Fig.~\ref{fig-scattering}c). Then
the multi-particle amplitude for any noninteracting evolution is given by
a sum of $2^M$ determinants $2M\times 2M$ composed of $4M\times 2M$ elements
of the scattering matrix spanned by the input and output channels.
For notational convenience, we group these elements into $4M$ rows of $2M$ elements each
and denote these rows $v_1, \ldots, v_{4M}$. Then the multi-particle amplitude (multiplied
by the factor $2^M$ for convenience) is
\begin{equation}
D_{2,2}(v_1, \ldots, v_{4M})=\sum_{i_k=0,1, \atop k=1,\ldots,M}
\det\begin{bmatrix} 
v_{2i_1+1} \\ v_{2i_1+2} \\ v_{2i_2+5} \\ v_{2i_2+6} \\ \vdots \\
v_{2i_M+4M-3} \\ v_{2i_M+4M-2}
\end{bmatrix}\, ,
\label{def-1}
\end{equation}
where the matrix in the right-hand side is composed of the corresponding rows $v_i$.
For each pair of rows, two pairs of vectors $v_i$ are considered 
(for the first pair of rows, either $v_1$ and $v_2$ or $v_3$ and $v_4$, and so on).
In linear algebra, the same function may be identified with the exterior product of two-forms
of rank two:
\begin{multline}
D_{2,2}(v_1, \ldots, v_{4M})
= (v_1 \wedge v_2 + v_3 \wedge v_4) \\
\wedge (v_5 \wedge v_6 + v_7 \wedge v_8) \wedge \ldots \\
\wedge (v_{4M-3} \wedge v_{4M-2} + v_{4M-1} \wedge v_{4M})\, ,
\label{wedge-1}
\end{multline}
where $\wedge$ denotes exterior product (see, e.g., Ref.~\cite{winitski-2010} or
other textbooks).

I explicitly show that the above function is computationally \#P hard
by a reduction of a $N$-dimensional permanent to this
function at $M=N^2$. This proves that, modulo a polynomial overhead, the
considered multi-particle amplitude is at least as computationally difficult as
as a permanent, which, in turn, is known to be \#P complete \cite{valiant-1979}.
The details of the proof are presented in Section \ref{sec:Proof}.

An alternative proof of the \#P-hardness of this function was communicated to me by
L.~Gurvits \cite{gurvits-private} based on a relation to mixed discriminants
\cite{barvinok-1997,gurvits-2005,gurvits-2007}. 
This proof is outlined in Section \ref{sec:MD}.

Section \ref{sec:Discussion} contains a brief discussion of the result,
including some simple generalizations and a possible extension to
approximate computations.

Finally, the Appendix contains an explicit form of the construction of the
proof of Section~\ref{sec:Proof} for the $N=3$ case.

\begin{figure}
\centerline{\includegraphics[width=.45\textwidth]{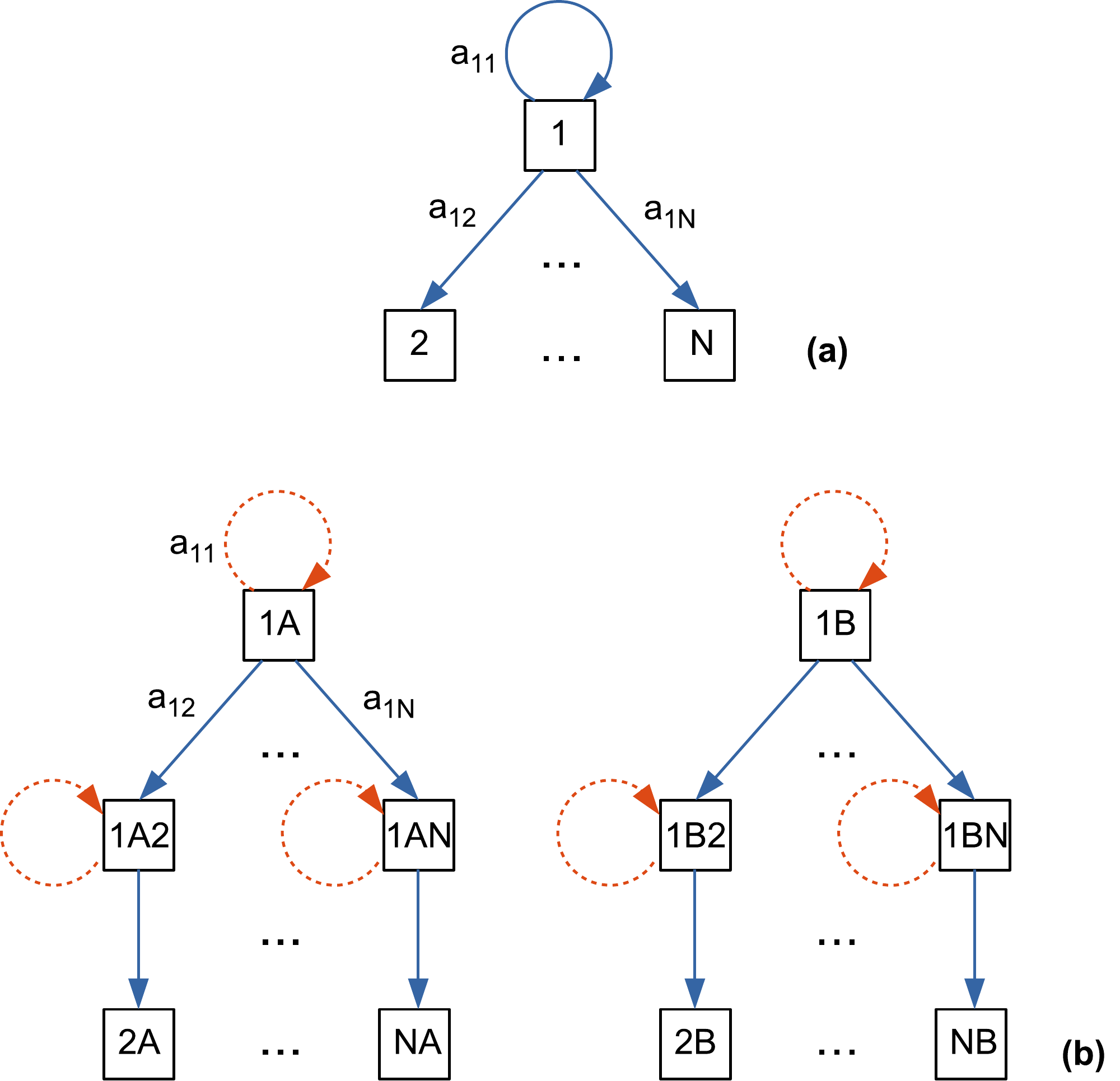}}
\caption{The construction of the two-color directed graph for the function 
(\ref{def-1})-(\ref{wedge-1})
[panel (b)] from the graph for a permanent [panel (a)]. The weights of the
directed edges are labeled as $a_{ij}$. The unlabeled edges have weight one.
The nodes of the two-color directed graph are grouped in pairs: 1A together with 1B,
1A2 together with 1B2, and so on (every node labeled with letter A is paired
with the corresponding node with letter B). Within each pair, the color of the
outgoing edges in the cycle cover must be unique.}
\label{fig-graphs}
\end{figure}

\section{Proof of \#P hardness using permanents}
\label{sec:Proof}

The proof can be most easily formulated in terms of graphs. The permanent of a matrix
$A$ of dimension $N$ with coefficients $a_{ij}$ is defined by Eq.~(\ref{perm-1}).
We can think of $A$ as a weighted adjacency matrix for a graph with $N$ nodes,
so that $a_{ij}$ is the weight attributed to the edge directed from $i$ to $j$.
In this representation, $\perm (A)$ can be thought of as the sum of products of
weights over all cycle covers of this directed graph \cite{papadimitriou-1994}.

A similar representation is possible for the function (\ref{def-1})-(\ref{wedge-1}). 
Namely, consider a directed graph with $2M$ nodes and edges colored in two colors
(dubbed ``color 1'' and ``color 2'' below and shown as solid 
and dashed arrows in the figures). Let vectors $v_1$ and $v_2$ contain
weights attributed to color-1 edges, originating from nodes 1 and 2, respectively, 
vectors $v_3$ and $v_4$ contain weights attributed 
to color-2 edges, originating from the same nodes 1 and 2, respectively, and so on.
Generally, vectors $v_{4i-3}$ and $v_{4i-2}$ correspond to color-1 edges originated
from nodes $2i-1$ and $2i$, respectively, and vectors $v_{4i-1}$ and $v_{4i}$
correspond to their color-2 counterparts. Then $D_{2,2}(v_1, \ldots, v_{4M})$
can be viewed as the sum of products of weights, multiplied by the corresponding signs,
over all cycle covers of this directed graphs under the constraint that, 
for each pair of nodes (1,2), (3,4), \ldots, ($2M-1$, $2M$), the cycle cover uses
the same color for edges originating from the two nodes in the pair. The
sign factor is determined as the parity of the total number of cycles.

The idea of the proof is to construct, for each directed graph for a matrix permanent,
a two-color directed graph for the function (\ref{def-1})-(\ref{wedge-1}), so that the loop covers
are in one-to-one correspondence in the two graphs and produce the same weights.
In order to cancel the sign factors, we double the number of nodes: the nodes in the
two-color graph will be denoted as ``A nodes'' and ``B nodes'', and the edges will only
connect nodes of the same type. At the same time, the coloring scheme will be used in
such a way as to constrain the cycle cover of A nodes to exactly repeat the cycle
cover of B nodes, so that the sign factor cancels out.

The construction of such a two-color directed graph is shown in Fig.~\ref{fig-graphs}.
Without loss of generality, we consider node 1. It has $N$ outgoing edges (to the same
node and to the $N-1$ other nodes). This node and the outgoing edges are replaced by
$2N$ nodes ($N$ A nodes and $N$ B nodes) and the corresponding outgoing edges as shown
in the figure. The A nodes are paired with the corresponding B nodes (in the figure, 
node 1A forms a pair with node 1B, node 1A2 with node 1B2, etc.).

This construction is repeated for each node of the original directed graph for the 
matrix permanent. As a result, the two-color directed graph for the function
(\ref{def-1})-(\ref{wedge-1}) contains $2N^2$ nodes.

On inspection, the constraint of the cycle covers in the two-color directed graph
guarantees that the cycles in the A nodes exactly reproduce the cycles in the corresponding
B nodes. This cancels out the sign factor. At the same time, the product of the weights
of the edges reproduces the product of the edges in the corresponding cycle cover
of the directed graph for the permanent. This proves that the function 
(\ref{def-1})-(\ref{wedge-1})
calculated for the constructed two-color directed graph equals the permanent of the
matrix $(a_{ij})$.

An example of the $3\times 3$ matrix is presented in Appendix.

Since the computation of the permanent of a matrix with integer elements is \#P-complete 
\cite{valiant-1979}, this proves that the computation of the function $D_{2,2}(v_1, \ldots, v_{4M})$ 
for integer-valued vectors $v_i$ is \#P-hard (a computation of the permanent can be reduced
to this function with a polynomial time overhead).

\begingroup
\squeezetable
\begin{table}[t]
\begin{tabular}{c||cc|cc|cc|cc|cc|cc|cc|cc|cc|}
 \hline\hline
 $v_1$ & $a_{11}$ &&&&&&&&&&&&&&&&&\\ 
 $v_2$ && 1 &&&&&&&&&&&&&&&&\\
 \hline 
 $v_3$ &&&&&&& $a_{12}$ && $a_{13}$ &&&&&&&&&\\
 $v_4$ &&&&&&&& 1 && 1 &&&&&&&&\\
 \hline\hline
 $v_5$ &&& $a_{22}$ &&&&&&&&&&&&&&&\\
 $v_6$ &&&& 1 &&&&&&&&&&&&&&\\
 \hline 
 $v_7$ &&&&&&&&&&& $a_{21}$ && $a_{23}$ &&&&&\\
 $v_8$ &&&&&&&&&&&& 1 && 1 &&&&\\
 \hline\hline
 $v_9$ &&&&& $a_{33}$ &&&&&&&&&&&&&\\
 $v_{10}$ &&&&&& 1 &&&&&&&&&&&&\\
 \hline 
 $v_{11}$ &&&&&&&&&&&&&&& $a_{31}$ && $a_{32}$ &\\
 $v_{12}$ &&&&&&&&&&&&&&&& 1 && 1 \\
 \hline\hline
 $v_{13}$ &&&&&&& 1 &&&&&&&&&&&\\
 $v_{14}$ &&&&&&&& 1 &&&&&&&&&&\\
 \hline 
 $v_{15}$ &&& 1 &&&&&&&&&&&&&&&\\
 $v_{16}$ &&&& 1 &&&&&&&&&&&&&&\\
 \hline\hline
 $v_{17}$ &&&&&&&&& 1 &&&&&&&&&\\
 $v_{18}$ &&&&&&&&&& 1 &&&&&&&&\\
 \hline 
 $v_{19}$ &&&&& 1 &&&&&&&&&&&&&\\
 $v_{20}$ &&&&&& 1 &&&&&&&&&&&&\\
 \hline\hline
 $v_{21}$ &&&&&&&&&&& 1 &&&&&&&\\
 $v_{22}$ &&&&&&&&&&&& 1 &&&&&&\\
 \hline 
 $v_{23}$ & 1 &&&&&&&&&&&&&&&&&\\
 $v_{24}$ && 1 &&&&&&&&&&&&&&&&\\
 \hline\hline
 $v_{25}$ &&&&&&&&&&&&& 1 &&&&&\\
 $v_{26}$ &&&&&&&&&&&&&& 1 &&&&\\
 \hline 
 $v_{27}$ &&&&& 1 &&&&&&&&&&&&&\\
 $v_{28}$ &&&&&& 1 &&&&&&&&&&&&\\
 \hline\hline
 $v_{29}$ &&&&&&&&&&&&&&& 1 &&&\\
 $v_{30}$ &&&&&&&&&&&&&&&& 1 &&\\
 \hline 
 $v_{31}$ & 1 &&&&&&&&&&&&&&&&&\\
 $v_{32}$ && 1 &&&&&&&&&&&&&&&&\\
 \hline\hline
 $v_{33}$ &&&&&&&&&&&&&&&&& 1 &\\
 $v_{34}$ &&&&&&&&&&&&&&&&&& 1 \\
 \hline 
 $v_{35}$ &&& 1 &&&&&&&&&&&&&&&\\
 $v_{36}$ &&&& 1 &&&&&&&&&&&&&&\\
 \hline\hline
 \end{tabular}
\caption{Vectors $v_i$ for the identity (\ref{app-1}).
Each of these vectors has 18 components (grouped
in pairs for better visualization). 
Empty spaces denote zeros.}
\label{app-table}
\end{table}
\endgroup

\section{Alternative proof using mixed discriminants}
\label{sec:MD}

I am grateful to L.~Gurvits \cite{gurvits-private} for bringing to my attention the 
following alternative proof using the theory of mixed discriminants
\cite{barvinok-1997,gurvits-2005, gurvits-2007}. Namely, Theorem 3.4 of \cite{gurvits-2005} 
states the \#P-hardness of computing the mixed discriminant $D(A_1, \dots, A_M)$
of rank-2 real symmetric positive semidefinite matrices $A_i=x_{i,0} x^*_{i,0} + x_{i,1} x^*_{i,1}$. 
If we introduce the $4M$ vectors in the $(M{+}M)$-dimensional space $v_{4i-3}=x_{i,0} \oplus 0$, 
$v_{4i-2}=0 \oplus x_{i,0}$,  $v_{4i-1}=x_{i,1} \oplus 0$, 
$v_{4i}=0 \oplus x_{i,1}$ (with $i=1,\ldots,M$), then one can verify that the
mixed discriminant can be expressed in terms of the exterior product (\ref{wedge-1}) as
\begin{equation}
D(A_1, \dots, A_M) = (-1)^{M(M-1)/2} D_{2,2}(v_1, \ldots, v_{4M})
\end{equation}
(the above relation follows, e.g., from Lemma 5.2.1 of Ref.~\onlinecite{bapat-1997}).
This proves the \#P-hardness of the exterior product (\ref{wedge-1}).

\section{Discussion}
\label{sec:Discussion}

The above proof admits several simple generalizations and corollaries.

First, the above relation between the permanent and the function (\ref{def-1})-(\ref{wedge-1})
holds for coefficients in any field or, even more generally, in any commutative ring.

Second, one can generalize the function (\ref{wedge-1}) to an exterior product
of $k$-forms of rank $\le r$ in $kM$-dimensional linear space,
\begin{equation}
D_{k,r}(\omega_1,\ldots \omega_M) = \omega_1 \wedge \ldots \wedge \omega_k\, ,
\label{wedge-2}
\end{equation}
where all $\omega_i$ are $k$-forms of rank $\le r$. The function $D_{2,2}$ is 
the simplest nontrivial example of this construction. Moreover, for any
$k\ge 2$ and $r \ge 2$, the function $D_{k,r}$ includes $D_{2,2}$ as
a particular case. This implies that the more general function
$D_{k,r}$ is also \#P-hard (for $k \ge 2$ and $r \ge 2$). On the other hand, 
for $k=1$ or $r=1$, the function $D_{k,r}$ is the determinant and is computable 
in polynomial time.

The proof above shows that the exact computation of the function
(\ref{def-1})-(\ref{wedge-1})
is at least as difficult as the exact computation of the permanent. Yet another
interesting question is an approximate computation. While the permanent of a
matrix with positive entries admits an efficient (randomized) approximate calculation
in polynomial time \cite{jerrum-2004}, even an approximate calculation 
(up to a multiplicative factor)
of a permanent in the general case is believed to be exponentially hard 
\cite{aaronson-2011,gurvits-2005,aaronson-2014}. 
If this is indeed the case, the worst-case running time of an approximate calculation
of the function (\ref{def-1})-(\ref{wedge-1}) should also be exponential.

Like in the Boson-Sampling case, the \#P hardness of the scattering
amplitudes does not automatically imply the quantum supremacy of the
proposed Fermion-Sampling setup. A proof (or a refutation)
of such a quantum supremacy would go beyond the scope of this paper and presents a
serious challenge, similarly to the Boson-Sampling case \cite{aaronson-2013}.

\section{Acknowledgments}

I thank A.~G.~Abanov, V.~Fock, P.~Roushan, and A.~Shen for discussions. 
I am grateful to L.~Gurvits for comments on the
manuscript and for bringing to my attention an alternative proof (Section \ref{sec:MD}).
This work was supported by the Swiss National Foundation through the NCCR QSIT.

\section{Appendix}

The simplest nontrivial case that illustrates the construction
shown in Fig.~\ref{fig-graphs} is the $3\times 3$ matrix. In that case,
we have the identity
\begin{equation}
\perm\begin{pmatrix}
  a_{11} & a_{12} & a_{13} \cr
  a_{21} & a_{22} & a_{23} \cr
  a_{31} & a_{32} & a_{33} \cr
\end{pmatrix} =
D_{2,2} (v_1,\ldots,v_{36})\, ,
\label{app-1}
\end{equation}
where the vectors $v_1$, \ldots, $v_{36}$ are listed in Table~\ref{app-table}.



\end{document}